\documentclass[structabstract]{aa}  
\usepackage{natbib}
\usepackage{graphicx}
\usepackage{txfonts}
\usepackage[usenames]{color}

\begin{document}
   \title{NGC~6340: an old S0 galaxy with a young polar disc}

   \subtitle{Clues from morphology, internal kinematics and stellar populations}

   \author{I. V. Chilingarian\inst{1,2}
          \and
          A. P. Novikova\inst{2}
         \and
          V. Cayatte\inst{3}
           \and
          F. Combes\inst{1}
          \and
          P. Di Matteo\inst{4}
          \and
          A. V. Zasov\inst{2}
}

   \offprints{Igor Chilingarian \email{igor.chilingarian@obspm.fr}}

   \institute{Observatoire de Paris-Meudon, LERMA, UMR~8112, 61 Av. de l'Observatoire, Paris, 75014, France
         \and
              Sternberg Astronomical Institute, Moscow State University, 13 Universitetski prospect, 119992, Moscow, Russia
         \and
             Observatoire de Paris-Meudon, LUTH, UMR~8102, 5 pl. Jules Janssen, Meudon, 92195, France
         \and
             Observatoire de Paris-Meudon, GEPI, UMR~8111, 5 pl. Jules Janssen, Meudon, 92195, France
}

   \date{Received April 10, 2009; accepted April 30, 2009; in original
form January 20, 2009}
   \authorrunning{Chilingarian et al.}
   \titlerunning{NGC~6340: Kinematics and Stellar Populations}

 
  \abstract
   {Lenticular galaxies are believed to form by a combination of environmental effects and 
secular evolution.}
   {We study the nearby disc-dominated S0 galaxy NGC~6340
photometrically and spectroscopically to understand the mechanisms of S0
formation and evolution in groups.}
   {We use SDSS images to build colour maps and light profile of NGC~6340
which we decompose using a three-component model including S\'ersic and two
exponential profiles. We also use Spitzer Space Telescope archival near-infrared
images to study the morphology of regions containing warm interstellar medium 
and dust. Then, we re-process and re-analyse deep long-slit
spectroscopic data for NGC~6340 applying novel sky subtraction technique and
recover its stellar and gas kinematics, distribution of age and
metallicity with the NBursts full spectral fitting.}
   {We obtain the profiles of internal kinematics, age, and metallicity
out to $>2$ half-light radii. The three structural components of NGC~6340
are found to have distinct kinematical and stellar population properties.
We see a kinematical misalignment between inner and outer regions of the
galaxy. We confirm the old metal-rich centre and a wrapped inner gaseous
polar disc ($r \sim 1$~kpc) having weak ongoing star formation, counter-rotating in
projection with respect to the stars. The central compact pseudo-bulge 
of NGC~6340 looks very similar to compact elliptical galaxies.}
   {In accordance with the results of numerical simulations, we conclude
that properties of NGC~6340 can be explained as the result of a major merger
of early-type and spiral galaxies which occurred about 12~Gyr ago. The
intermediate exponential structure might be a triaxial pseudo-bulge formed
by a past bar structure. The inner compact bulge could be the result of a
nuclear starburst triggered by the merger. The inner polar disc appeared
recently, 1/3--1/2~Gyr ago as a result of another minor merger or cold gas
accretion.}

   \keywords{evolution of galaxies --  galaxies: kinematics --
             galaxies: stellar populations -- galaxies: individual
(NGC~6340) }

   \maketitle
%

\section{Introduction}

According to the morphological classification of galaxies \citep{Hubble36},
lenticular, or S0 galaxies represent a transitional galaxy type between
ellipticals and spirals. The presence of massive stellar discs makes them
resemble spiral galaxies, although lenticulars have, in average, higher
bulge luminosities, lower contrast of spiral arms (if any) and  lower
H{\sc i} surface density, as well as very weak star formation. Global colour
properties put lenticular galaxies on the ``red sequence'' in the
colour--luminosity relation (e.g. \citealp{Strateva+01}).

Galaxy morphology is often connected to their environment: \citet{HH31} were
the first to point out the differences between field and cluster galaxy populations.
\citet{SB51} and \citet{GG72} have suggested that dynamical effects govern the
life of galaxies in clusters and groups: collisions of late-type galaxies should
produce early-type ones, and the ram pressure stripping by the intracluster
medium would effectively expel the ISM from gas-rich systems (see also
\citealp{QMB00}). \citet{Moore+96} proposed another mechanism of
the late-to-early type morphological transformation, namely ``gravitational
harassment'' or numerous interactions of a galaxy with other cluster members,
although not as catastrophic as major mergers.

\citet{LTC80} described a different scenario: star formation should strongly
deplete the gas in most spirals in a couple of Gyr resulting in the S0-like
appearance of galaxies, whereas present-day spirals must have been
experiencing external supply of gas, i.e. by accreting tidal debris, minor
mergers with gas-rich satellites, or cold gas from cosmic filaments.

Most proposed mechanisms of spiral-to-lenticular morphological
transformation result in a gas concentration in the central region of a
galaxy triggering a strong circumnuclear starburst, raising the average stellar
metallicity. 
This gas concentration should also
 decrease the age of the stellar population, if
the event occurred quite recently. Given that the peak of S0 formation must
have happened at $z \sim 0.4 - 0.5$, in a considerable fraction of
lenticulars the nuclear starburst should have occurred no more than 5~Gyr
ago, which is confirmed by recent observations revealing chemically- and
evolutionary-decoupled nuclei in many nearby lenticular galaxies
\citet{Silchenko06}.

Early-type disc galaxies have been intensively studied during last two
decades. One half of the initial sample of 48 early-type galaxies observed in
the course of the SAURON project (see Table~3 in \citealp{Emsellem+04}) are
classified as S0s or barred lenticulars. All of them but two were later
classified as fast rotators \citep{Emsellem+07}. Stellar populations of
early-type disc galaxies \citep{Silchenko06,Kuntschner+06,SGC06,Peletier+07}
exhibit great diversity of properties, possibly suggesting the importance of
environmental effects on their evolution. However, in most cases,
observations did not go beyond one half-light radius ($r_e$), i.e. providing
information only about bulge-dominated regions.

Here we present the studies of internal kinematics and stellar population
properties of the lenticular galaxy NGC~6340 out to $>2 r_e$, where the
observed stellar light is dominated by the outer regions of its disc.

NGC~6340 is an early-type disc galaxy, classified as S0-a in the HyperLeda
database\footnote{http://leda.univ-lyon1.fr} \citep{Paturel+03}. Being a
group member it does not have any close companions of comparable luminosity.
The imagery reveals a bright central concentration and low-contrast spiral
structure in the outer parts \citep{ZMKS08}. The bulge is quite bright
containing about a quarter of a total galaxy luminosity. \citet{Silchenko00}
discovered an old metal-rich nucleus and circumnuclear polar ring inside $r
= 0.5$~kpc (or 6~arcsec).
 She studied the morphology with the HST image, showing an inner dust 
lane \citep{CSdZM97,CS98}, and derived the velocity field with an integral-field
spectroscopy, of field of view $10 \times 16$~arcsec. In this paper, we derive 
kinematics in the long-slit mode out to 80~arcsec from the centre.

Throughout the rest of this paper we assume the following general properties
of NGC~6340: distance of 17~Mpc (assuming $H_0 = 73$~km~s$^{-1}$~Mpc$^{-1}$,
$v_r = 1230$~km~s$^{-1}$) corresponding to the spatial scale of
82~pc~arcsec$^{-1}$ and distance modulus $m - M = 31.15$~mag. The galaxy
position on the sky corresponds to 0.25, 0.18, 0.13, 0.10, and 0.07~mag of
Galactic extinction \citep{SFD98} in the $u$, $g$, $r$, $i$, and $z$ bands
respectively.

The paper is organized as follows: in the next Section we present the
surface photometry and analysis of light profiles of NGC~6340; Section~3
contains details about spectroscopic observations, data reduction and
analysis as well as the kinematical and stellar population properties of the
galaxy; in Section~4 we discuss the results obtained.

\section{Morphology and internal structure}

NGC~6340 is contained in the imaging footprint of the Sloan Digital Sky
Survey Data Release 6 (SDSS DR6, \citealp{SDSS_DR6}). The data were collected on
20/Sep/2001 using the 2.5~m SDSS telescope at the Apache Point Observatory
in the $u$, $g$, $r$, $i$, and $z$ photometric bands with the corresponding
atmosphere FWHM seeing quality of 1.3, 1.2, 1.0, 0.9, and 1.0~arcsec.

The data have been corrected for the atmosphere extinction and converted
into absolute fluxes and corresponding calibrated AB-magnitudes using
prescriptions available on the web-site of the SDSS
project\footnote{http://www.sdss.org/}. After having completed several tests
we concluded that the sky background subtraction could be done by a simple
subtraction of a constant level specified in the FITS-headers of the
corresponding data files: attempts of modelling the sky background with the
2-dimensional polynomial did not reveal any statistically significant
deviations from the flat level at a region of the frame containing the
galaxy.

\begin{figure}
\includegraphics[width=\hsize]{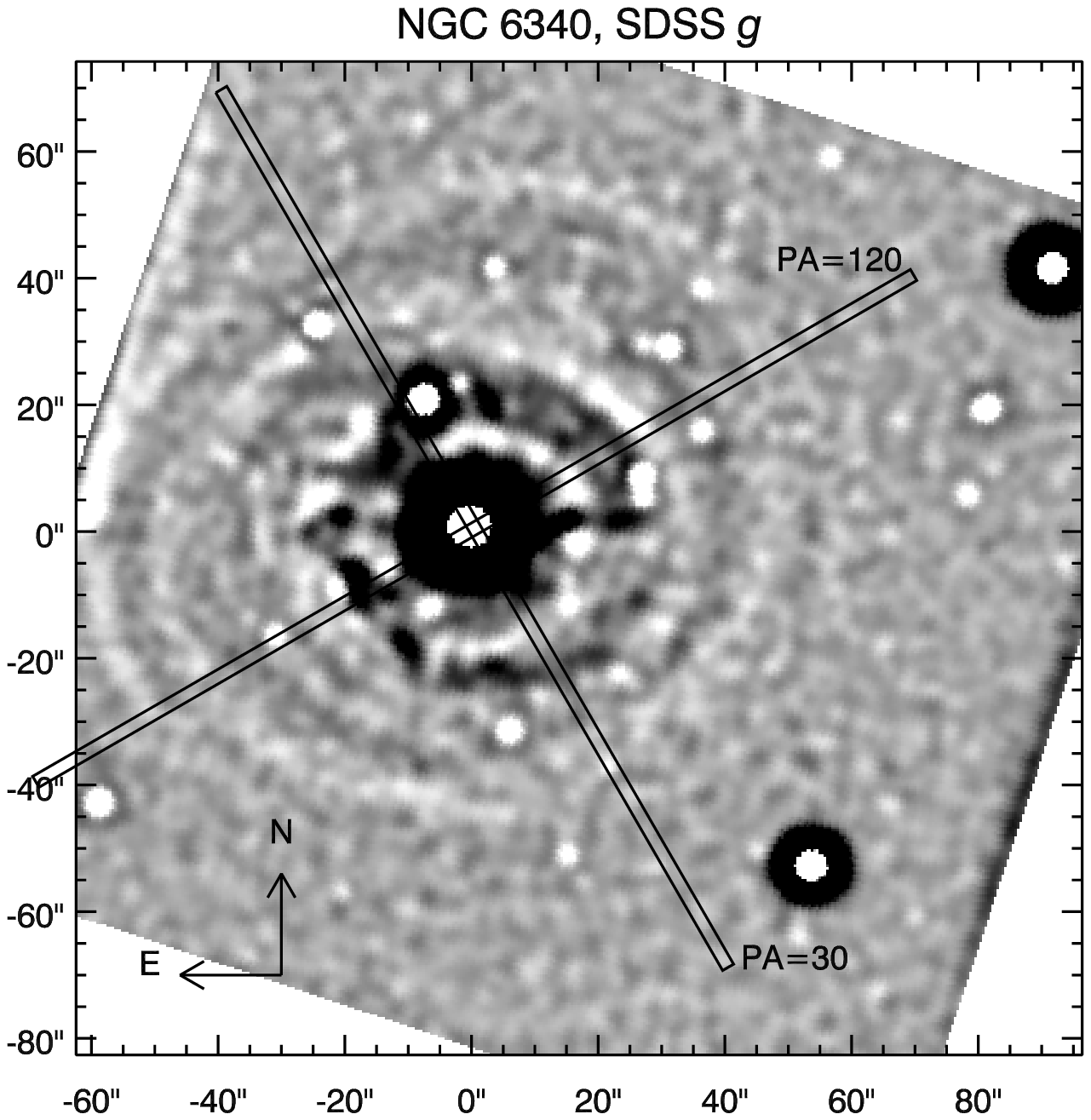}
\includegraphics[width=0.49\hsize]{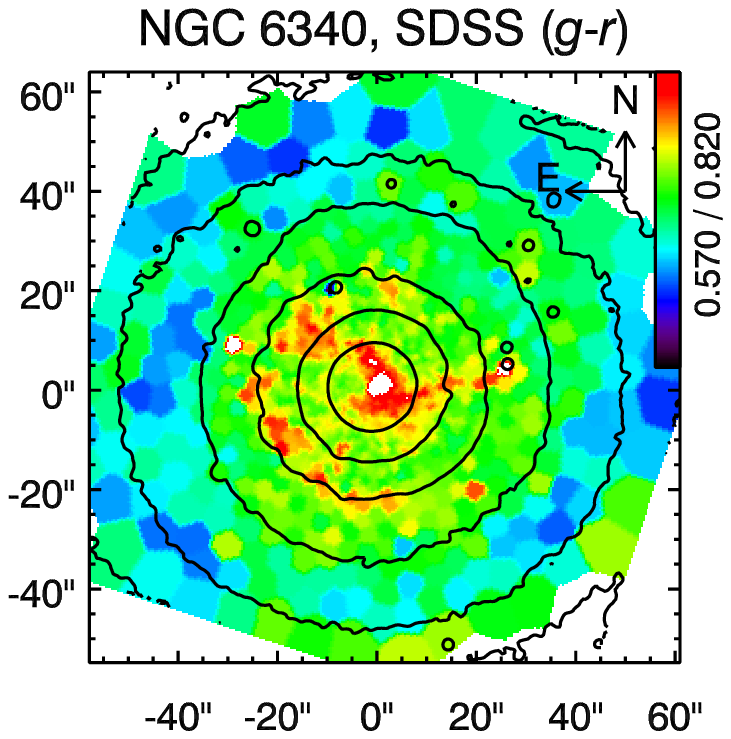}
\includegraphics[width=0.49\hsize]{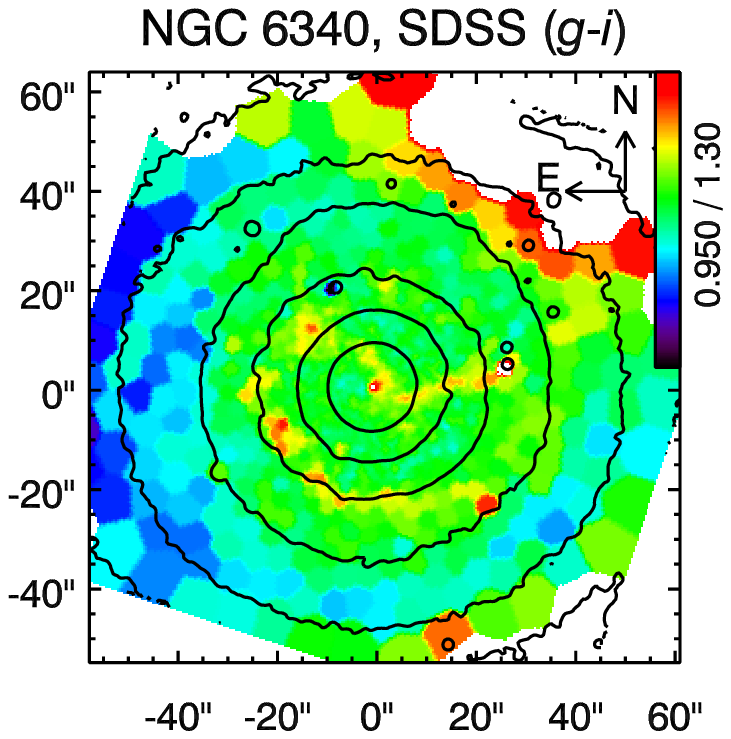}
\caption{The $g$-band unsharp-masked image of NGC~6340 with the
SCORPIO slit positions overplotted (top); $g - r$ (bottom left) and
$g - i$ (bottom right) colour maps derived from the SDSS images using
adaptive binning (see text). The bins having the areas exceeding 120
sq.~pixels were masked.\label{figcolmaps}}
\end{figure}

The unsharp-masked $g$ band image of NGC~6340 showing its low-contrast
shells or spiral arm fragments and dust lanes is displayed in the top panel
of Fig.~\ref{figcolmaps}. It was obtained by subtracting a Gaussian-convolved
image (FWHM$ = 5.2$~arcsec) from the original data and, again, convolving
the result with the two-dimensional Gaussian having FWHM$ = 2.8$~arcsec.
 There is no obvious winding sense for these potential spiral arms, and in
the original image, they look more like plateaus of emission, with a sharp
drop outside, characterising shells formed in mergers \citep{HQ88,DC86}. The
presence of shells all around the center, with random orientation are
typical of small companions accreted by an oblate potential \citep{DC86}.

We built the colour maps of NGC~6340 in order to study the distribution of
dust in it. To proceed with this we firstly convolved the images in all
photometric bands with the 2-dimensional circular Gaussians corresponding to
the squared difference between the corresponding atmosphere seeing and
1.8~arcsec. Then, we have applied the Voronoi adaptive 2D-binning using the
algorithm and software package described in \citet{CC03} to reach the target
signal-to-noise ratio of 80 per bin in the sky subtracted $r$-band image.
After that, the binning configurations were used for image tessellation in
all 5 bands. 

Two colour maps constructed in this fashion are presented in
Fig.~\ref{figcolmaps}. The dust lanes are clearly visible in the $g - i$ map.
Areas with weak ongoing star formation having bluer colours are apparent in
the $g - r$ map, for example, one 20--25~arcsec North-East of the galaxy
centre tracing a fragment of a spiral arm or shell. Other spiral arm fragments are
also noticeable in the outer parts of the presented colour maps. The red colour
gradient inwards is evident.

We have fit the elliptical isophotes into the images of NGC~6340 using the
algorithm described in \citet{Jedrzejewski87} and implemented as the {\sc
stsdas.analysis.isophote.ellipse} task in the {\sc iraf} data processing
environment. Prior to the isophote fitting, we created the masks to minimize
the influence of foreground stars and NGC~6340 globular clusters on the
obtained photometric information. We did this by using the iterative
procedure including the {\sc ellipse} task calls alternating with the
threshold-based detection of regions to be included in the mask in the
Gaussian-convolved residual images.

\begin{figure}
\includegraphics[width=\hsize]{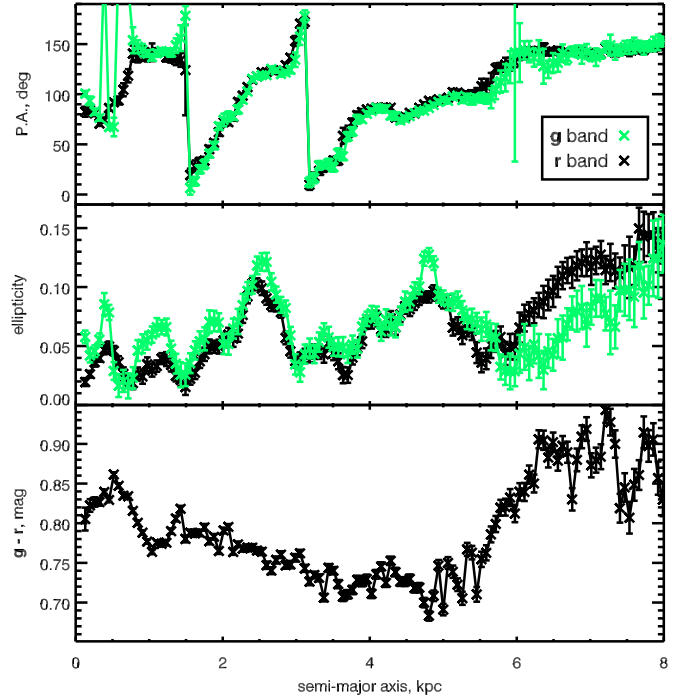}
\caption{Radial behaviour of the major axis positional angle (top),
ellipticity (middle) of the isophotes of NGC~6340 from the
SDSS $g$ and $r$-band images shown in green and black respectively. The $g -
r$ colour profile reconstructed from the isophote fitting is displayed in
the bottom panel.\label{figpaell}}
\end{figure}

In Fig.~\ref{figpaell} we present the radial behaviour of $e = 1 -
b/a$ and positional angle (two top panels) in $g$ and $r$ band and the $g -
r$ colour profile. The galaxy exhibits very round isophotes with the
ellipticity below 0.06 in the inner part ($R < 1.5$~kpc). The region between
$1.5 < R < 3.7$~kpc is affected by the presence of spiral arms or shells
traced by strong variations of positional angle. Then, the P.A. stays nearly
constant at $\sim$90~deg out to $R \sim 5.5$~kpc. In the outermost measured
regions of the galaxy the P.A. changes to 145~deg and remains constant from
$R \sim 6$~kpc while the ellipticity increases to 0.15. The outer isophotes
are slightly rounder in $g$ than in $r$. 

The colour profile (bottom panel in Fig.~\ref{figpaell}) comprises
several features: red central part, modest negative gradient at $1.5 < R <
5.5$~kpc, and redder flat outer part. Since the $r$ band profile is the
deepest among five, less affected by the effects of dust and spiral
arms than the $g$ band and by the two bright red stars North-West of the
galaxy than the $i$ band, we used it to perform further analysis.

We have performed the structural decomposition of the $r$-band light
profile. Using two-component decomposition including an inner
\citet{Sersic68} and an outer exponential components did not result in a
fitting having satisfactory quality: the procedure converged only if we
excluded the innermost region, at the same time systematic differences
between the model and the data were evident at radii beyond 40~arcsec. The
residuals were significant between 12 and 25~arcsec, where an excess of
light over an outer exponential disc is evident.

Therefore we decided to model the light distribution of NGC~6340 using
three-component model comprising the inner S\'ersic and two exponential
profiles. This modelling resulted in a good quality of fitting everywhere
from the galaxy centre out to 140~arcsec.

\begin{figure}
\centering
\includegraphics[width=\hsize]{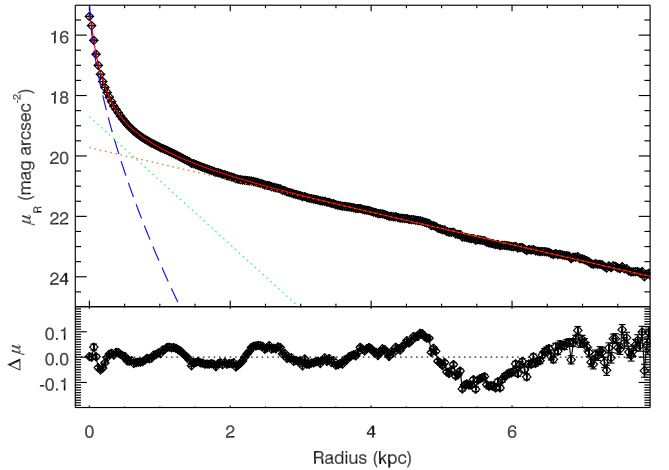}
\caption{The three-component NGC~6340 light profile decomposition. The top
panel displays the brightness profile shown with black diamonds, and the
three components represented by the blue dashed, green and red dotted lines
for inner S\'ersic, internal and external exponential profiles
correspondingly. The bottom panel shows the fitting residuals.
\label{figdecompos}}
\end{figure}

We used the nonlinear optimization using the Levenberg-Marquardt technique
of the 7 parameters of all three components simultaneously (S\'ersic profile
presented by its effective radius, effective surface brightness and $n$
index, two exponential profiles presented by their central surface
brightness values and exponential lengths). We fitted the data points at 
radii $0 \le R \le 140$~arcsec. In order to fit the central surface
brightness value, the model was convolved with the radially-averaged point
spread function determined empirically by measuring several individual stars
on a CCD frame in the regions close to the galaxy. 
Note that
a procedure similar to ours was applied to perform the light profile
decomposition of 2 barred lenticulars by \citet{EBGB03} and the first
evidences of nested exponential structures were given.

The procedure is very sensitive to the initial guess, therefore we had to
proceed as follows to find it:
\begin{itemize}
\item Fitting only the outer part of the profile ($R > 40$~arcsec) with the
single-component exponential model to get the parameters of the outer disc.
\item Fitting the two-component model of the profile ($R > 10$~arcsec)
fixing the outer disc parameters in order to get the initial guess for the
parameters of the inner disc.
\item The four parameters were fixed and the inner S\'ersic profile was fit.
\item The resulting set of 7 parameters was used as an initial guess to fit
the three-component model varying all 7 parameters.
\end{itemize}

The parameters of the best-fitting 3-component model of the $r$-band light
profile are presented in Table~\ref{tabdecomp}. 

\begin{table}
\caption{NGC~6340 $r$-band light profile decomposition using the 3-component
model. All value are corrected for the Galactic extinction. Absolute
magnitudes of the components are converted into the $B$ band using the
transformation from \citet{FSI95} for lenticular galaxies $B - r =
1.17$~mag.} 
\begin{tabular}{lccc}
\hline
\hline
& S\'ersic & 1st Disc & 2nd Disc \\
\hline 
$r_{e}$ kpc & 0.195 $\pm$ 0.003 & & \\
$n$  & 1.66 $\pm$ 0.04 & & \\
$\mu_{e}$ mag~arcsec$^{-2}$ & 17.99 $\pm$ 0.16 & & \\
$\mu_{0}$ mag~arcsec$^{-2}$ & & 18.46 $\pm$ 0.55 & 19.60 $\pm$ 0.15 \\
$d_{\mbox{exp}}$ kpc & & 0.49 $\pm$ 0.02 & 2.02 $\pm$ 0.02 \\
$\langle\mu\rangle_{e}$ mag~arcsec$^{-2}$ &  17.04 $\pm$ 0.16 & 19.59 $\pm$ 0.55 & 20.72 $\pm$ 0.15 \\
$M_{B}$~mag & $-16.88$ & $-17.47$ & $-19.41$ \\
\hline
\hline
\end{tabular}
\label{tabdecomp}
\end{table}

The outer large scale disc contains
80~per~cent of the total galaxy luminosity, 
with an exponential length of about $\sim$2~kpc and a central
surface brightness 20.8~mag~arcsec$^{-2}$ converted into the $B$-band.
The inner disc contains
13~per~cent of the luminosity, having a four
times smaller exponential length, however a higher $B$-band surface brightness
of 19.6~mag~arcsec$^{-2}$. The remaining 7~per~cent of the galaxy light are
coming from a very compact high-surface brightness ($\langle\mu\rangle_{e,B}
= 18.21$~mag~arcsec$^{-2}$) nuclear bulge with a half-light radius of only
about 0.2~kpc. The total luminosity of NGC~6340 recovered from our
3-component model corrected for the Galactic extinction and converted into
the $B$-band, $M_B = -19.66$~mag, places the galaxy into the class of
intermediate-luminosity lenticulars.

\begin{figure}
\includegraphics[width=\hsize]{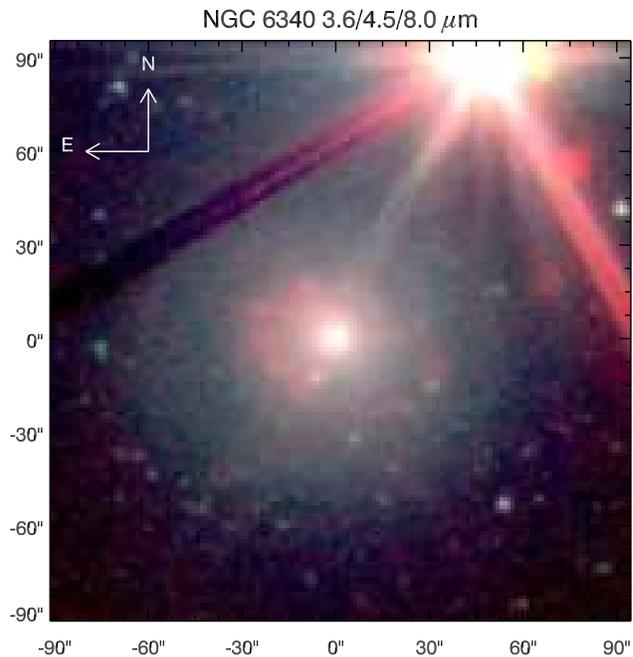}
\caption{False colour composite image of NGC~6340 constructed from 3.6
(blue), 4.5 (green), and 8.0 (red) data obtained with the Spitzer Space
Telescope.\label{fignir}}
\end{figure}

NGC~6340 was observed with the Spitzer Space Telescope in the frame of the
nearby galaxy survey. We accessed the fully calibrated 4 band near-infrared
data presented in \cite{PAFW04} through the {\sc leopard} tool available for
download at the Spitzer Space Telescope Archive. The galaxy images are
strongly affected by the bright star 100~arcsec North-West of NGC~6340.
However, qualitative analysis is possible from colour maps constructed from
images at 3.6 to 8~$\mu$m. In Fig.~\ref{fignir} we present the false colour
composite image constructed from NIR Spitzer data. The 8~$\mu$m data shown
in red, tracing the presence of poly-aromatic hydrocarbons in the regions
with ongoing star-formation are of particular interest. The galaxy exhibits
similar features to the optical colour maps, but they are more evident. We
see a structure resembling a one-arm spiral, which seems to be connected to
the peculiarities of emission-line kinematics presented and discussed below.

\section{Insights from deep long-slit spectroscopy}

\subsection{Observations, data reduction and analysis}

The spectroscopic observations of NGC~6340 were obtained in the course of
the observing project ``Discs of lenticular galaxies'' (P.I.: AZ) using the
SCORPIO universal spectrograph \citep{AM05} mounted at the prime focus of
the 6-m Bol'shoy Teleskop Azimutal'nyy (BTA) at the Special Astrophysical
Observatory of the Russian Academy of Sciences (SAO RAS). The long-slit
spectroscopic mode of SCORPIO using the VPHG2300G grating provides an
intermediate spectral resolution ($R \approx 2200$) in the wavelength range
covering blue-green spectral region ($4800 < \lambda < 5500$~\AA) with the
6~arcmin long 1.0~arcsec wide slit. The 2048$\times$2048 pixels EEV CCD42-40
detector was used binning the data by a factor of 2 along the slit resulting
in the spectral sampling of about 0.75~\AA~pix$^{-1}$ and spatial scale of
0.357~arcsec~pix$^{-1}$.

The kinematical analysis of these data for have already been published
\citep{ZMKS08} together with three other lenticular galaxies. The
observations we refer to were obtained during two observing runs in
May 2005 (P.A.=120~deg, t$_{exp}$=7200~s, seeing=3.5~arcsec) and July
2006 (P.A.=30~deg, t$_{exp}$=9644~s, seeing=1.6$\dots$2.5~arcsec). Poor
atmosphere transparency and cloudiness severely affected the first observing
run reducing the effective exposure time to $\sim 3000$~s.

The positional angles of the slits were chosen based on the major axis $P.A.
= 120$~deg reported in the UGC catalogue \citep{Nilson73}. The slit
positions are shown in Fig.~\ref{figcolmaps} (top panel) on top of the
unsharp-masked image of the galaxy. The $P.A.= 30$~deg dataset has very
high signal-to-noise ratio exceeding 150 per spectral element per slit pixel
in the centre of the galaxy.

Due to the flexure of the telescope and spectrograph described in
\citet{Moiseev08}, arc-lines and internal flat field calibration frames were
taken at night time during observations. Additional calibration included
spectrophotometric standard stars and high-resolution twilight spectra which
we used to measure and then take into account variations of the
spectrograph's line-spread-function (LSF).

We reduced the data in the {\sc itt idl} package derived from the data
reduction package developed at SAO RAS. The primary data reduction steps
comprising bias subtraction, flat fielding, removing cosmic-ray hits using
Laplacian filtering \citep{vanDokkum01} were applied to all science and
calibration frames. Then, we built the wavelength solution by identifying
arc lines and fitting their positions with the two-dimensional polynomial of
the 3rd order in the both dimensions, along and across dispersion and
linearized the spectra. The obtained wavelength solution had fitting
residuals of about 0.08~\AA~RMS mostly due to the statistical errors of the
determined arc line positions. We did not increase the power of the
polynomial surface, because the fitting procedure becomes unstable and very
sensitive to the positions of individual (faint) arc lines, and at the same
time we have a technique to take into account the systemic errors during the
data analysis. The error frames were computed using the photon statistics
and processed through the same reduction steps as the data.

After that, we binned the linearized twilight spectra obtained during the
corresponding observing runs using 64 equal 16-pixels wide intervals along
the slit to increase the signal-to-noise ratio to several hundred per
spectral element. We fitted the high-resolution ($R=10000$) solar spectrum from
the ELODIE.3.1 \citep{PSKLB07} stellar library against these twilight
spectra in five wavelength segments overlapping by 20~per~cent covering the
spectral range of the SCORPIO setup using the penalized pixel fitting
procedure by \citet{CE04}. The radial velocities deviating from zero
obtained from the fitting indicated the systemic errors of the wavelength
solution mapped over the field of view and wavelength range of the
spectrograph, while velocity dispersion and higher-order moments of the
Gauss-Hermite parametrization $h_3$ and $h_4$ \citep{vdMF93} gave information
about the spectral resolution and deviations of the SCORPIO's LSF from
Gaussian. The coefficients of the LSF parametrization along the slit are 
smoothed using splines.

All details regarding the spatial, spectral and time variations of the
SCORPIO's LSF will be given in Novikova \& Chilingarian (in prep.), here we
give a short summary. (1) The systemic errors of the wavelength solution
change smoothly from -15~km~s$^{-1}$ to zero from blue to red end of the
covered spectral domain with little variations along the slit of on order of
5~km~s$^{-1}$. The behaviour is very well reproduced between the observing
runs (within 1--2~km~s$^{-1}$) and is probably connected to imprecise
tabulated wavelengths of the blended arc lines in the blue part or/and
insufficient power of the polynomial used to fit the 2D wavelength solution.
(2) Spectral resolution ($\sigma_{\mbox{inst}}$), $h_3$, and $h_4$ demonstrate
significant variations along the slit: in its central part the LSF is very
close to Gaussian with $\sigma_{\mbox{inst}} = 65$~km~s$^{-1}$ at all
wavelengths, only $h_3$ remains modestly negative, whereas it degrades toward
outer slit regions, by about 40~per~cent at the lower part of the CCD-frame
(slit positions between 0 and 1~arcmin) and half of this at the upper part.
The deviations from Gaussian also become very important, reaching $h_3 =
-0.13$ and $h_4 = 0.08$. (3) Behaviour of $\sigma_{\mbox{inst}}$, $h_3$, and
$h_4$ in the red part of the wavelength range ($\lambda > 5250$\AA) depend of
the focusing of the spectral camera, but remain stable during the observing
run once the focus position has been set. This explains the importance of
obtaining twilight spectra during every observing run.

The mapping of the SCORPIO's LSF is essential for the precise
sky subtraction. Since we did not have separate sky exposures we had to
construct the model of the air-glow night sky emission based on the spectra
from the peripheral regions of the slit which were not contaminated by the
galaxy's light. The LSF exhibits important variations and spectral
resolution degrades toward outer parts of the slit, therefore we proposed the
following procedure for creating the night sky model.
\begin{enumerate}
\item We create a high signal-to-noise night sky spectrum by co-adding
spectra over large regions in the outer parts of the slit
\item The LSF properties and its variations along the wavelength for this
co-added night sky spectrum are obtained by fitting the Solar spectrum as
explained above using the twilight spectrum assembled from the same regions
of the slit
\item In several slightly overlapping wavelength intervals (usually, 5 or
6), where the wavelength-dependent LSF variations can be neglected, we
perform the correction of the LSF in the co-added sky spectrum bringing its
shape to one determined at a given position along the slit, hence
constructing a model sky spectrum at a given position, using the
mathematical properties of convolution in the Fourier space as follows:
\begin{equation}
f(x, \lambda) = F^{-1}(F(f(\mbox{sky}, \lambda))
\frac{F(\mathcal{L}(x))}{F(\mathcal{L}(\mbox{sky})}),
\end{equation}
\noindent where $f(x, \lambda)$ denotes a sky spectrum at the position $x$
along the slit having its parametrized LSF $\mathcal{L}(x)$; $f(\mbox{sky},
\lambda)$ is the co-added night sky spectrum with the LSF
$\mathcal{L}(\mbox{sky})$, and $F$, $F^{-1}$ are for the direct and inverse
Fourier transforms respectively.
\end{enumerate}

The night sky spectrum at every slit position created in this way has the
LSF corresponding to one determined by fitting the twilight
spectra. The usage of parametrized LSF instead of Fourier-transformed
original twilight spectra at a given slit position is required to achieve a
certain level of regularization and avoid noise amplifications at some
frequencies which may be caused by the Fourier-based signal transformation
technique described above. 

The resulting model of the night sky emission was subtracted from the
spectra. Then, we performed the flux calibration using the observations of
the spectrophotometric standards.

The application of this technique to the NGC~6340 data produced excellent
results: for the $P.A.=30$~deg dataset we are able to reliably subtract the
night sky emission even in the regions of the galaxy having a $B$-band
surface brightness as low as 25~mag~arcsec$^{-2}$.

The spectra of NGC~6340 were binned adaptively along the slit by starting
from the photometric centre and co-adding consequent pixels outwards until a
given target signal-to-noise ratio had been reached. This approach allowed
us to analyse the data even in the periphery of the galaxy where the surface
brightness was quite low by degrading the spatial resolution of the data.
All further spectral data analysis was applied independently to every
spatial bin.

We fitted high-resolution {\sc pegase.hr} \citep{LeBorgne+04} simple
stellar population (SSP) models against the observational data using the
{\sc NBursts} full spectral fitting technique \citep{CPSA07,CPSK07}. The
models were computed using the \cite{Salpeter55} stellar initial mass
function and the high-resolution stellar library ELODIE.3.1 \citep{PSKLB07}.
The fitting algorithm works as follows: (1) a grid of SSP spectra with a
fixed set of ages (nearly logarithmically spaced from 20~Myr to 18~Gyr) and
metallicities (from $-$2.0 to $+$0.5~dex) is convolved with the
wavelength-dependent instrumental response of SCORPIO as explained in
Section~4.1 of \cite{CPSA07}; (2) a non-linear least square fitting against
an observed spectrum is done for a template picked up from the pre-convolved
SSP grid using 2D-spline interpolation on $\log t$ and $Z$, broadened
according to the line-of-sight velocity distribution (LOSVD) parametrized by
$v$, $\sigma$, $h_3$, and $h_4$ and multiplied pixel-by-pixel by the
$n^{\rm{th}}$ order Legendre polynomial, resulting in $n + 7$ parameters
determined by the non-linear fitting. The penalization based on the values
of high-order Gauss-Hermite moments is applied to the $\chi^2$ as explained
in \citet{CE04} in order to bias the LOSVD towards pure Gaussian in case of
low signal-to-noise ratio and/or insufficient spectral sampling. In our case
the reliable determinations of $h_3$ and $h_4$ are achievable only in the inner
10~arcsec from the galaxy centre.

For the data analysis presented hereafter we adopted the 13$^{\rm{th}}$ order
multiplicative continuum and no additive continuum. The needs for
multiplicative continuum and possible side-effects have been presented and
analysed in detail in Appendix~A2.3 of \citet{CPSA07}, Appendix~B1 of
\citet{Chilingarian+08}, and \citet{Koleva+08}, where it was also shown that
the {\sc NBursts} full spectral fitting produces consistent stellar
population parameters with those derived from the measurements of the Lick
indices \citep{Worthey94}, being several times more precise.

To avoid possible biases of the stellar population parameters caused by the
contamination of the spectra by emission lines, we have excluded the
20~\AA-wide regions around H$\beta$ ($\lambda = 4861$~\AA), [O{\sc iii}]
($\lambda = 4959, 5007$~\AA) and [N{\sc i}] ($\lambda = 5199$~\AA) emission
lines redshifted accordingly to the mean radial velocity of the galaxy as
well as the Hg{\sc i} ($\lambda = 4561$~\AA) line originating from the light
pollution. \citet{Chilingarian09} demonstrated that H$\beta$ contains
20~per~cent of the age-sensitive information at maximum when using {\sc
NBursts} technique in the spectral range similar to ours, therefore
excluding it from the fitting neither biases age estimates (see also
Appendix~A2 in \citealp{CPSA07} and Appendix~B in
\citealp{Chilingarian+08}), nor degrades significantly the quality of the
age determination. For our study it is also important that: (1) velocity
dispersions can be precisely determined at down to 1/3--1/2 of the spectral
resolution (i.e. 25--30~km~s$^{-1}$ for our data) and these measurements
remain unbiased (see e.g. \citealp{KBCP07} and Section~2 in
\citealp{CCB08}); (2) luminosity-weighted values of ages and metallicities
are insensitive to the $\alpha$/Fe ratios of the populations being fit
\citep{Chilingarian+08,Koleva+08} even though {\sc pegase.hr} models at
solar and moderately subsolar metallicities are representative of
[$\alpha$/Fe] = 0.0~dex. Here we fit the spectra using single-SSP models,
therefore the returned age and metallicity estimates are SSP-equivalent.

Uncertainties of our estimates of stellar kinematics are at least twice
lower than those published in \citep{ZMKS08} mostly due to better matching
of the NGC~6340 by SSP models compared to empirical stellar templates and to
more precise sky subtraction in the outer regions of a galaxy. In addition,
a precise modelling of the stellar continuum enabled us to extract the
ionised gas kinematics from very faint emission lines left in the
fitting residuals.

\subsection{Stellar kinematics}

\begin{figure}
\includegraphics[width=\hsize]{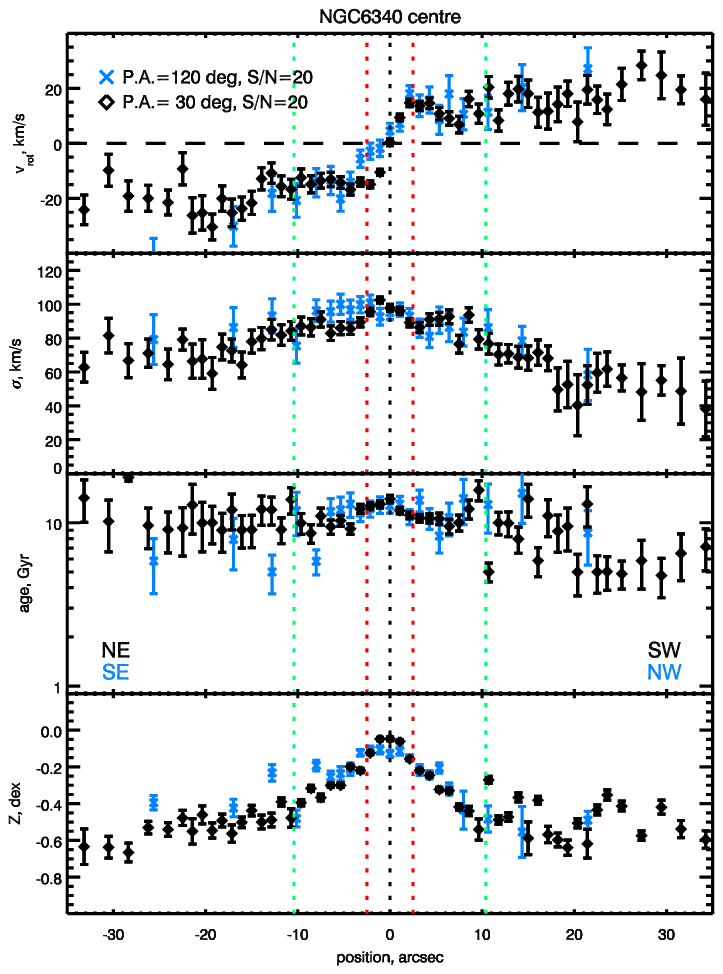}
\includegraphics[width=\hsize]{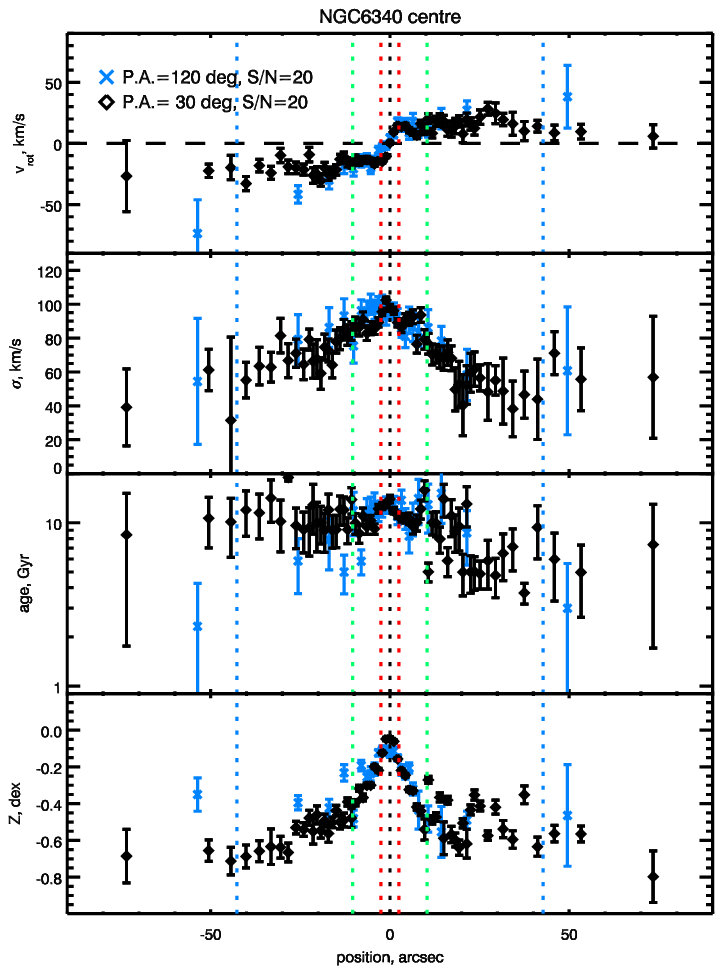}
\caption{Kinematics and stellar populations of NGC~6340. Top and bottom
groups of panels display the zoomed-in inner region and the whole profiles
out to 1.5~arcmin from the centre. Top to bottom in each group: radial 
velocities, velocity dispersions, ages and metallicities.
Blue and black data points are for the $P.A.=120$ and $P.A.=30$~deg profiles
respectively. Colored vertical dashed lines denote half-light radii of the
three substructures in the light profile.
\label{figkinstpop}}
\end{figure}

The obtained profiles of radial velocities and velocity dispersion are
presented in Fig.~\ref{figkinstpop} (top two panels of each group of four).
The kinematical counterparts of the photometrically detected structural
components described in the previous section are clearly seen.

The rotation is immediately evident in both kinematical profiles suggesting
that initially selected slit positions did not correspond to the galaxy
major and minor axes. We qualitatively estimated the parameters of
orientation of the outer and intermediate structural components of NGC~6340
by deprojecting the kinematical profiles following the usual approach
for a thin disc.
Then, if the galaxy velocity field can be
represented with a pure disc rotation, the deprojected kinematical profiles
should coincide in case the systemic velocity ($v_0$), major axis positional 
angle ($P.A._{\mbox{major}}$), and inclination ($i$)
values are set correctly. Or, the other way round, the galaxy orientation
parameters and its systemic velocity can be determined assuming a model of
pure disc rotation by reaching the best agreement between the deprojected
profiles.

In case of NGC~6340 due to its almost face-on orientation, this technique
does not provide good precision. Nevertheless, even the qualitative
estimates of the orientation turn to be very different in the outer and inner
regions of the galaxy. The inner region ($R < 20$~arcsec) dominated by the
inner exponential structure in the light profile corresponds to
$P.A._{\mbox{major}} \approx 70$~deg, whereas in the outer parts 
($R > 25$~arcsec) it changes to $P.A._{\mbox{major}} \approx 100$~deg. The
inclination in both cases is between $20 < i < 25$~deg. Although, the
statistical errors of the estimates are as large as 5--7~deg, the kinematical
misalignment between the two substructures is evident.

We reach the flat part of the rotation curve at about $R = 40$~arcsec,
corresponding to 3.2~kpc. 

The absence of gaseous radial velocity profile and the low inclination of
the disc makes the estimate of circular velocity not too reliable.
Nevertheless, we estimated its approximate value from the available data.
Indeed, the line-of-sight velocity difference at $R = \pm 50$~arcsec, where
the velocity profile flattens, is $130 - 140$~km~s$^{-1}$ (see
Fig.~\ref{figkinstpop}). It nearly coincides with about 140~km~s$^{-1}$
obtained by \cite{Bottema93} for P.A.$=$130~deg and the same radial
interval, but a little higher than $\approx 100$~km~s$^{-1}$ within $R = \pm
40$~arcsec obtained from the same data with a different data processing
technique \citep{ZMKS08}. We accept the radial component of the velocity of
rotation to be close to 70~km~s$^{-1}$, which corresponds to the full
$v_{\mbox{rot}} \approx 185 \pm 20$~km~s$^{-1}$ after correction for $i =
20-25$~deg. For the line-of-sight velocity dispersion close to
50~km~s$^{-1}$ at these radii ($R \approx 1.5 d_{exp}$) and the ratio
$\sigma_z / \sigma_r \approx 0.7$ the circular velocity corrected for the
asymmetric drift is $v_c = 200\pm 30$~km~s$^{-1}$. The main contribution
into uncertainties is given by the low, poorly-determined disc inclination.
It is worth comparing this value with the total corrected observed line
width of the H{\sc i} line $W_{\mbox{H\sc{i}}} = 195$~km~s$^{-1}$
\citep{SHGK05}, which, after being divided by $2\sin i $ corresponds to $v_c
= 260\pm 30$~km~s$^{-1}$. This value exceeds by 30--40~km~s$^{-1}$ the
circular rotation obtained for the stellar disk, which may be explained if
we accept that H{\sc i} is concentrated in the inner 10~arcsec. Indeed, the
observed LOSVD within 10 arcsec for the emission lines is much higher than
for the stellar velocity of rotation reaching about 100~km~s$^{-1}$ (see
Fig.~\ref{figkinem}).

Having the maximal circular velocity of about 200~km~s$^{-1}$ and the
luminosity $M_B = -19.7$~mag, NGC~6340 nicely fits to the Tully--Fisher
(\citeyear{TF77}) relation for the lenticular systems, although the latter is
characterized by significant dispersion \citep{BAM06}. Hence, this galaxy
possesses a quite normal mass-to-light ratio for the lenticular galaxies of
similar luminosities. 

The velocity dispersion profiles have a little central bump $\sigma_0 =
105$~km~s$^{-1}$ corresponding to the inner compact pseudo-bulge and a plateau
having $\sigma_{\mbox{in}} = 90$~km~s$^{-1}$ at $R < 10$~arcsec
corresponding to the inner region of the inner exponential structure. Then
the values smoothly decrease to $\sim 65$~km~s$^{-1}$ in the large-scale
disc going down to $\sim 50$~km~s$^{-1}$ at $2 r_e$ of the outer disc
(5.8~kpc). 

The observed line-of-sight velocity dispersion of stellar components
of the disc of NGC~6340 along two P.A.s was recently compared with those
expected for the marginally stable disc using the method of numerical
modelling of the live collisionless disc embedded into the rigid
pseudo-isothermal halo \citep{ZMKS08}, where it was found that the disc of
this galaxy, unlike the discs of many spiral galaxies, is significantly
overheated, that is the observed velocity dispersion exceeds the minimal
values needed for the disc to be marginally stable to gravitational and
bending perturbations.  It supports the idea that this galaxy has
experienced a major merger or several minor mergers in the past.

The $h_4$ coefficient (not shown) demonstrates symmetric behaviour raising
from 0.03 in the centre to the maximal value of 0.08 at $R = 5$~arcsec and
then smoothly decreasing to zero at $R = 10 \dots 12$~arcsec. $h_3$ displays
modestly positive values at the eastern part of the galaxy (in both
datasets) starting from 0 at the centre, reaching the maximum of 0.06 about
$5 \ldots 8$~arcsec east of the centre then going down to zero at $R >
12$~arcsec, at the same time staying at zero level west of the centre. This
strange behaviour evidently reflects the presence of two overlapping
structures with different kinematical properties and relative luminosities.

\subsection{Stellar populations}

The profiles of the SSP-equivalent age and metallicity measurements are
presented in Fig.~\ref{figkinstpop} (two lower panels in each group). 

The metallicity profiles exhibit clear three-component structure: (1)
central plateau with the constant level metallicity $-0.02 \pm 0.005$~dex and
a size corresponding to the inner compact pseudo-bulge; (2) inner sharp
exponential gradient (i.e. linear in the plots since the metallicity is
already presented in logarithmic units of dex) in the region corresponding
to the inner exponential structure with a break at $\sim12$~arcsec; (3)
outer weak exponential gradient, corresponding to the large-scale disc.

We have measured the parameters of metallicity gradients in the 2$^{\rm{nd}}$ and
3$^{\rm{rd}}$
regions by fitting two-side linear functions in the regions $2 < R < 12$ and
$20 < R < 70$~arcsec of the $P.A.=30$~deg dataset and, hence, obtained the
extrapolated central values and gradient slopes per kpc and per
characteristic exponential length of a given structure. The central
extrapolated metallicities measured in this manner were found to be
$[\mbox{Fe/H}]_{0,\mbox{in}} = -0.08 \pm 0.01$~dex and
$[\mbox{Fe/H}]_{0,\mbox{out}} = -0.37 \pm 0.03$~dex for the inner and outer
discs respectively. The corresponding gradient slopes per kpc and per
exponential length are $d[\mbox{Fe/H}]_{\mbox{in}}/dR = -0.47 \pm
0.01$~dex/kpc$ = -0.23 \pm 0.01$~dex/exp.l$_{\mbox{in}}$ and
$d[\mbox{Fe/H}]_{\mbox{out}}/dR = -0.055 \pm 0.010$~dex/kpc$ = -0.11 \pm
0.02$~dex/exp.l$_{\mbox{out}}$.

The age profile also traces the presence of three structural components, but
at the same time it displays the asymmetric behaviour in the inner region
evident in the measurements derived from the $P.A.=30$~deg dataset. The
North-East part ($t = 13.5 \pm 0.8~Gyr$) is $\sim 3$~Gyr older than
South-West ($t = 10.5 \pm 0.6$~Gyr), although at $R \sim 5 \dots 7$~arcsec
at both sides the age becomes similar ($t = 10$~Gyr). The outer disc also
exhibits similar asymmetry in the age distribution, being generally younger
($t = 6 \dots 9$~Gyr) than the inner region of NGC~6340. This is probably
connected to the presence of an thin inclined disc containing dust and a
small amount of young stellar population (see below).

We would like emphasize here that the excellent quality of spectroscopic
data we analyse in this work results in the statistical errors of the
stellar population parameters significantly beyond the quality of the
evolutionary synthesis models. Therefore, the absolute values of ages and
metallicities should not be trusted at the error levels provided. They may
change if we, for instance, replace ELODIE.3.1 with another stellar library,
or use different evolutionary tracks when running the {\sc pegase.hr} code.
Studies of all these effects are far beyond the scope of this paper.
However, the differential behaviour of recovered stellar population
properties (i.e. gradients, asymmetries, differences of age/metallicity
between different structural components) is reliable.

Our results qualitatively agree with \citet{Silchenko00} arguing for
the presence of a chemically decoupled nucleus in NGC~6340. The metallicity
difference we obtain between the nucleus and at $R=5$~arcsec, $\Delta Z
\approx 0.30$~dex coincides with the measurements by \citet{Silchenko00}
based on the iron line-strength index. Nuclear and bulge age estimates
obtained from more recent analysis of the same data \citep{Silchenko06}
given their large uncertainties are also in agreement with our measurements.

\subsection{Kinematics of ionized gas}

We have analysed the emission line kinematics of NGC~6340 by fitting
single-component Gaussians into the residuals of the stellar population
fitting at the positions corresponding to H$\beta$ and [O{\sc iii}]
($\lambda = 5007$~\AA). The resulting positions and line widths were
corrected correspondingly to the spectrograph's LSF at a given slit position
and wavelength. The emission line fluxes obtained from the best-fitting as 
products of line intensities and widths were used to calculate the [O{\sc
iii}]/H$\beta$ emission line ratio.

\begin{figure}
\includegraphics[width=\hsize]{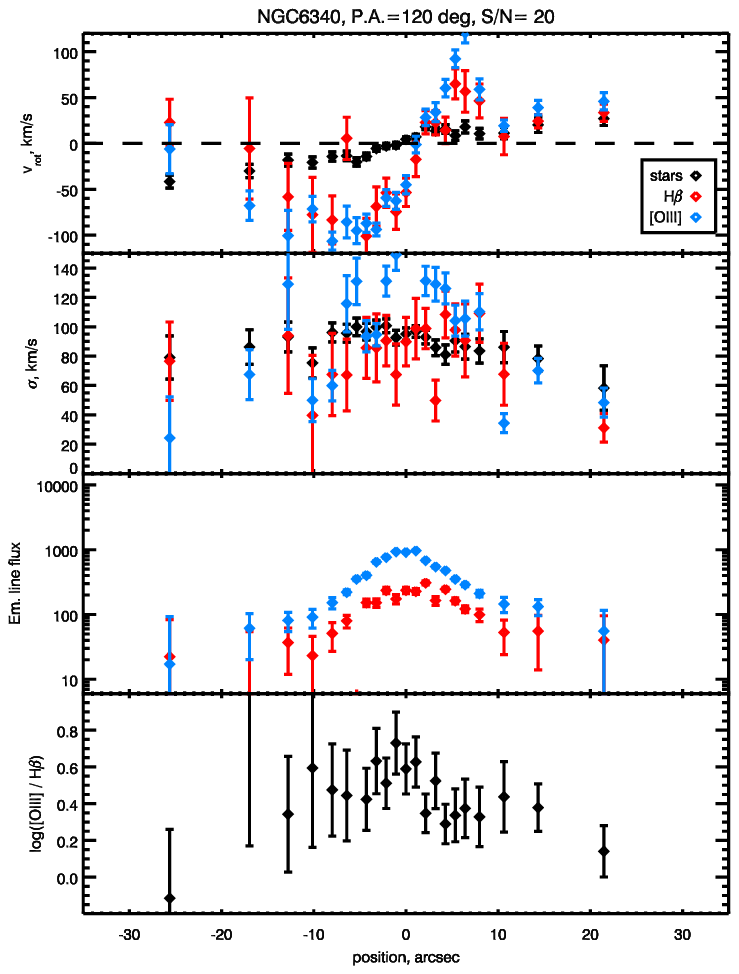}
\includegraphics[width=\hsize]{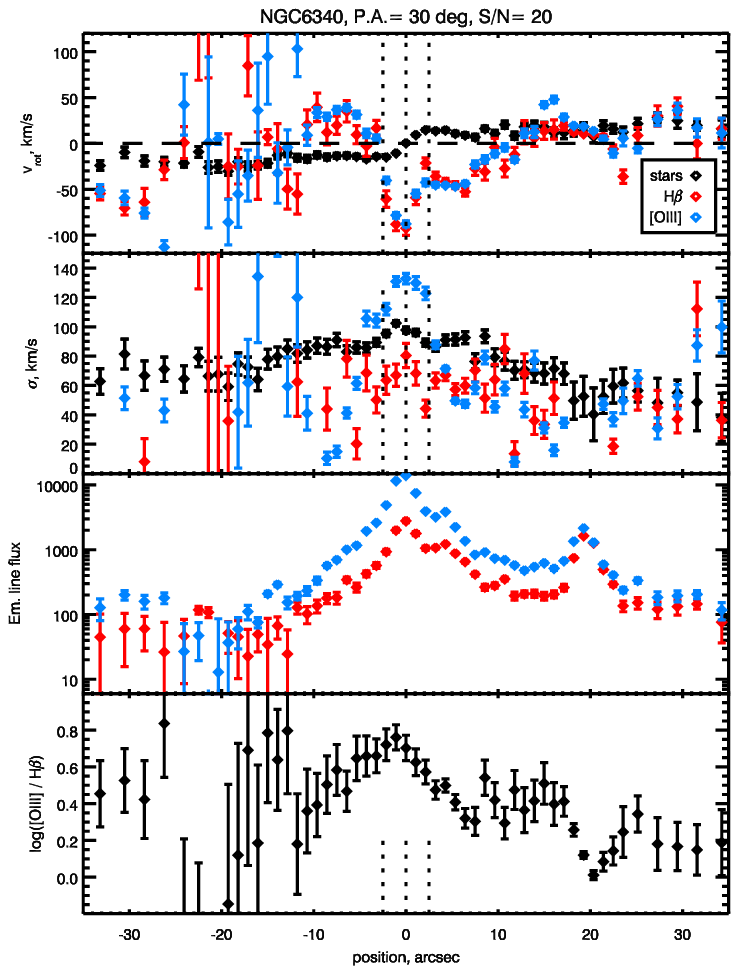}
\caption{Kinematics of ionized gas compared to stars and properties of emission
lines from the analysis of the H$\beta$ (red) and [O{\sc iii}] ($\lambda =
5007$~\AA, blue) lines in the spectra of NGC~6340 with the best-fitting
stellar population models subtracted. The panels show (top to bottom):
radial velocities, velocity dispersion, line fluxes in arbitrary units,
logarithm of the emission line ratio. Top and bottom groups of panels are for the
$P.A=120$ and $P.A.=30$~deg datasets respectively.
\label{figkinem}}
\end{figure}

The ionized gas in the inner region of NGC~6340 exhibits fast rotation at $5
< R < 10$~arcsec, reaching $100$~km~s$^{-1}$ in projection in the
$P.A.=120$~deg dataset. At the same time it counter-rotates to the stars in
the $P.A.=30$~deg dataset arguing for a presence of the highly inclined
disc-like structure. The central parts of both profiles are very irregular:
central values in the $P.A.=120$~deg slice differ from the stellar velocities
by about $-60$~km~s$^{-1}$ which was obvious yet from the Fig.~10 of
\citet{Silchenko00}. The $P.A.=30$~deg profile explains this discrepancy: in
the circumnuclear region we see a sharp ``drop'' of the radial velocity
($-80$~km~s$^{-1}$) slightly (1~arcsec) offset from the photometric centre of the
galaxy to NE. Interestingly, the intensity of the forbidden nitrogen line
[N{\sc ii}] in this direction is higher than in the opposite which is
clearly seen by the asymmetric elongated toward bottom-left contours in
Fig.~7 of \citet{Silchenko00}.

Applying the kinematical deprojection technique to determine the orientation
of this disc is quite problematic because of the central region affected by
the ``drop''. Nevertheless, the approximate parameters of this structure
are: $P.A._{\mbox{major}} \approx -30$~deg, i.e. different by $-100$~deg from the
kinematical major axis of the inner exponential profile; inclination is
between 40 and 60~deg depending on which radii from the centre are used to
combine the profiles probably suggesting the strong warp of the observed
structure.

In the region corresponding to the ``drop'' of the radial velocity, the
[O{\sc iii}] emission lines exhibit very strong asymmetry and its width
reaches $\sigma \sim 150$~km~s$^{-1}$. The emission lines are stronger
in the western part of the galaxy than in the east, which is evident from
the line flux plots (2$^{\rm{nd}}$ bottom panels) in Fig.~\ref{figkinem}.

\section{Discussion}

\subsection{Structural properties of NGC~6340}

In the previous section we have shown that the main stellar body of NGC~6340
comprises three well-defined structural components having different
properties of their stellar populations. 

The fast-rotating outer disc with high $v_c/\sigma \approx 3.5$ is
reminiscent of stellar discs in intermediate-luminosity spiral galaxies,
being at the same time modestly older. \citet{KvdKF05} claimed that the high
value of the circular velocity of NGC~6340 by \citet{Bottema93} was 
overestimated, however, we confirm that earlier result.

The age profile at $10 < R < 40$~arcsec exhibits statistically
significant ``jumps'' to lower values than in the surrounding regions with
their positions correlated to the loci of dust lanes in the colour maps.
Given the asymmetric distribution of dust and absence of a regular
spiral pattern in NGC~6340 with fragments of spiral arms or shells observed
instead, the asymmetric loci of the ``jumps'' with respect to the galaxy
centre are easy to understand.

The inner exponential structure exhibiting old metal-rich stellar
populations is supported mostly by random motions, given the deprojected
maximal circular velocity of $45 \pm 10$~km~s$^{-1}$ (assuming $i = 25$~deg)
reached at $R \sim 6$~arcsec, hence resulting in $v_c/\sigma \approx 0.5$.

\begin{figure}
\includegraphics[width=\hsize]{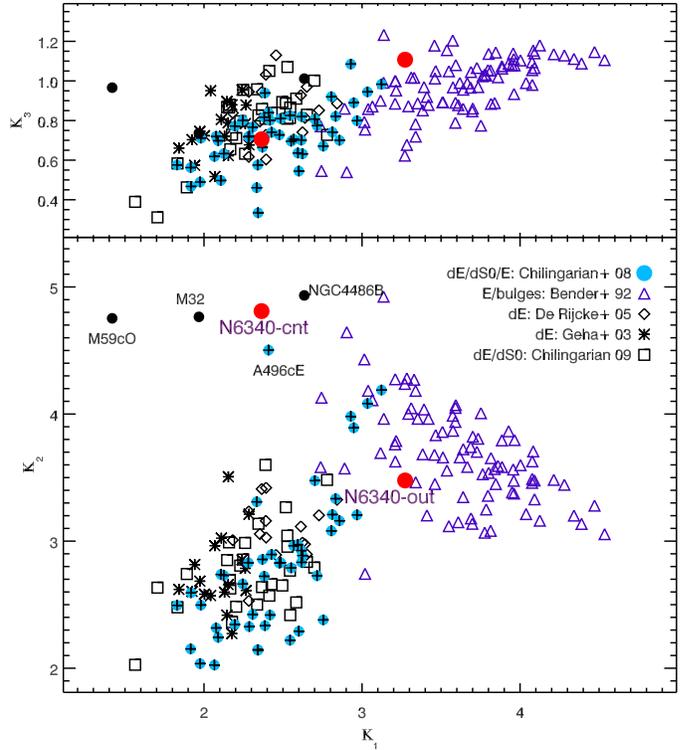}
\caption{$\kappa$-space view \citep{BBF92} of the Fundamental Plane. The
positions of the structural components of NGC~6340 are shown with filled red
circles and labeled as ``cnt'', ``out'' for the central compact
pseudo-bulge, and a superposition of the inner exponential component
and outer disc correspondingly. See the text for the sources of 
data.\label{figFP}}
\end{figure}

In Fig.~\ref{figFP}, we present the two projections of the Fundamental Plane
\citep[FP]{DD87} redefined in $\kappa$-space \citep{BBF92}, where $\kappa_1$
is related to the logarithm of the total mass, $\kappa_2$ is proportional to
the $(M/L) I_e^3$, hence measuring ``compactness'', and $\kappa_3$ is
connected to the logarithm of the dynamical mass-to-light ratio. On the
$\kappa_2$~vs.~$\kappa_1$ plot (the plane's ``face-on view''), the
boomerang-shaped area often interpreted as two distinct regions is occupied
by dwarf \citep{GGvdM03,deRijcke+05,Chilingarian+08,Chilingarian09} and intermediate
luminosity and giant early-type galaxies \citep{BBF92}.  There is an
extension of the sequence of giant galaxies towards the upper-left corner of
the plot by very rare compact elliptical (cE) galaxies with M59cO
\citep{CM08} being the most extreme case. Filled red circles display the
compact central pseudo-bulge and a superposition of intermediate and outer
exponential structures.
Since the FP is defined for random motion supported virialized stellar
systems, as a first-order approximation we have included the rotational
kinetic energy as $\sigma^2 + 0.5 v_c^2$ into the total energetic balance
for the inner exponential structure and large-scale disc of NGC~6340.

From Fig.~\ref{figFP} it is evident that the central compact pseudo-bulge of
NGC~6340 falls into the locus of cE galaxies. These objects are believed to
form through the tidal stripping of intermediate-mass disky progenitors
\citep{BCDG01,Chilingarian+07} and always found in the vicinities of massive
galaxies claimed to be responsible for the tidal stripping (Chilingarian et
al. in prep.) However, the numerical simulations still experience
difficulties in reproducing the dramatic increase of the central stellar
density required to form objects which can be observationally classified as
cEs. An example of an object initially having the stellar surface density
comparable to one in the cE galaxies is provided by the central pseudo-bulge
of NGC~6340. Then, if NGC~6340 had fallen onto a massive galaxy (e.g.
cluster cD) a compact elliptical galaxy could have been formed by the severe
tidal stripping of its extended components. Stellar population in the centre
of NGC~6340 is very similar to those observed in known cE galaxies
\citep{SGCG06b,Chilingarian+07}. In the inner region of NGC~6340,
dominated by the compact pseudo-bulge ($r=0.195$~kpc), we roughly estimate the mass
contribution of the two outer exponential components by integrating their
stellar mass profiles derived from the light profiles and stellar
mass-to-light ratios. Since the velocity dispersion of the intermediate
exponential structure is close to that in the galaxy centre, we expect its
thickness (assuming oblate morphology) to be comparable to the half-light
radius of the compact pseudo-bulge, thus, simple comparison of integrated
light profiles should provide a good approximation of their contributions to
the mid-plane gravitational potential. Our estimate of the stellar mass
contribution by the intermediate exponential disc to the total galaxy mass
within a half-light radius of the inner compact pseudo-bulge to be 
of 14--16~per~cent at maximum. Therefore, its removal would affect the central stellar
velocity dispersion only by 8--9~per~cent and will not change significantly
the position of the compact pseudo-bulge on the FP. The mass contribution 
from the outer large-scale disc will be below 2~per~cent.

\subsection{Inner polar disc in NGC~6340}

We confirm the existence of an inclined gaseous disc mentioned as a
``starforming polar ring'' in \citet{Silchenko00}. Given its orientation
obtained from the kinematical data and asymmetry in the profiles of NGC~6340
stellar age and emission line fluxes, we conclude that it is rotating
counter-clockwise with the SW part above the plane of the main stellar disc
of NGC~6340. Its plane is inclined by 40--65~deg with the respect to
the large scale stellar disc. If we consider that the polar disc contains
dusty ISM observed in the colour maps, and a little amount of young stars,
it will explain why the NE part of the galaxy looks older: the old
population in the SW part is partially obscured by the dusty young disc in
front of it. On the other hand, the main old stellar disc and pseudo-bulge
are also expected to contain significant quantities of dust
\citep{Driver+07} even lacking the current star formation, which will
explain why the emission line intensities in the part of the young
star-forming polar disc behind the main stellar disc are lower. 

At radii beyond 12~arcsec the ionised gas co-rotates with the stars
proving that we observe an \emph{inner} polar disc. Origin of inner polar
discs still remains a matter of debate. These structures are thought to form
by the gas settling onto one of the principal planes of a triaxial bulge
\citep{CPCB03,CCPB07} or a bar. In case of a large-scale gaseous disc
counter-rotating to the stars, the inner polar ring may be formed secularly:
gas will be captured onto highly inclined stable orbits on its way to the
galactic centre. However, in case of co-rotating gaseous and stellar
components, as we observe in NGC~6340, the externally supplied gas is needed
(see e.g. \citet{SA04} for a discussion).

The formation of large-scale (outer) and inner polar discs by accretion
from a companion has been simulated by \citet{BC03}, and from cosmic
filaments by \citet{MMS06} and \citet{Brooks+09}. When the accreted gas
encounters already pre-existing gas in the main galaxy, it collides and
dissipates, quickly getting aligned with the main gaseous disc through
differential precession, particularly if the accretion was not exactly polar
which seems the case for NGC~6340. However, when there is no pre-existing
gas, the accreted gas may remain on quasi-stable orbits for a long time.

Quite recently, an order of a few hundreds Myr ago, NGC~6340 probably
experienced a minor merger and accreted a low-mass gas-rich satellite. This
event created a structure presently observed both, in kinematical profiles
and direct images, as a counter-rotating inclined gaseous disc with a low
ongoing star formation and somewhat irregular lopsided structure of dust
lanes clearly revealed in the colour maps. This polar disc with a major axis
aligned approximately along east-west direction, contains low-contrast
spiral arms visible on the composite-colour image of the galaxy out to
$\sim$50~arcsec from the centre. The non-axisymmetric potential of the
large-scale exponential bulge of NGC~6340 drives the ISM of the star-forming
disc to the circumnuclear region following a complex trajectory, which we
probably see as a complex structure emitting at 8~$\mu$m at the Spitzer
Space Telescope images (Fig.~\ref{fignir}). The regions where it crosses the
slit at $P.A.=30$~deg are observed as a slightly offset from the centre
($\sim$1~arcsec to NE) deep negative peak of the ionized gas radial velocity
and the secondary positive peak at 16~arcsec SW. We notice that the emission
line ratio $\log ($[O{\sc iii}]/H$\beta)$ exhibits local maxima at the
corresponding regions reaching 0.8 in the (quasi)central peak, uniquely
arguing for the shockwave ionisation \citep{BPT81}. Additional spectral data
covering H$\alpha$ and [N{\sc ii}] are required for making further
conclusions about the ionisation mechanism.

Interestingly, the mentioned secondary peak of the gas radial velocity has
neither any visible counterparts in the intensity profiles of emission
lines, nor in the colour maps. At the same time, there is a prominent
secondary maximum of emission line intensities 19~arcsec SW of the galactic
centre ($P.A.=30$~deg slit). It corresponds to the locus of one of the dust
lanes in the colour maps and to the local minima of the [O{\sc
iii}]/H$\beta$ line ratio and velocity dispersions of both, stars and gas.
The luminosity-weighted age in this region is lower and the stellar
population is more metal-rich (Fig.~\ref{figkinstpop}). We explain this by
the superposition of old and young stellar populations from the pseudo-bulge
and the star-forming disc: even a low mass fraction of the dynamically cold
but young stellar population in the disc will have a large luminosity
fraction, biasing the estimations of kinematical and stellar population
parameters.

\subsection{Origin and evolution of NGC~6340}

As it was argued above, NGC~6340 is probably a result of a major
merger of two galaxies having unequal masses and/or different morphologies,
which happened over 12~Gyr ago and induced a major star formation event.
This hypothesis is supported by the internal structure of the galaxy,
distributions of its stellar content, and significant disc
overheating.

The central region of NGC~6340 corresponding to the innermost S\'ersic
component of the brightness profile with $n=1.6$, i.e. pseudo-bulge, and
having metal-rich stellar population ($[\mbox{Fe/H}]=-0.02$~dex) with no
radial gradients, was probably created by the strong star formation event
triggered by a major merger. Two super-massive black holes would cause very
fast dynamical relaxation resulting in complete mixing of the stars on a
short timescale which will wash out any possible structures in the metallicity
distribution. This process must have finished a very long time ago, because
the presently observed stellar population exhibits old age.

Then, we should interpret the intermediate exponential brightness
profile component as an exponential bulge, possibly having triaxial
structure and potential responsible for ``unusual'' kinematical parameters
derived from the deprojection of the kinematical profiles assuming disc
rotation. The properties of the large-scale outer disc suggest that the
masses and/or morphologies of the merged galaxies must have been different.
The outer disc exhibits modest internal velocity dispersion and the stellar
population younger than the inner region of the galaxy, which probably means
that the star formation in it had been continuing much longer and,
therefore, it was not so strongly affected by the process of interaction.
Due to its almost face-on orientation we are not able to measure reliably
the $v_c/\sigma$ parameter in order to conclude whether or not the disc was
strongly dynamically heated.

\begin{figure}
\includegraphics[width=\hsize]{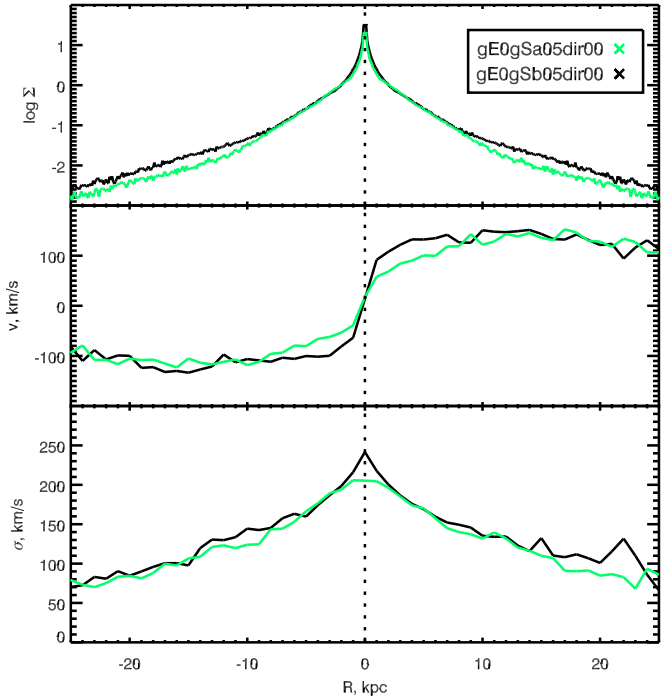}
\caption{Profiles of the surface density (top), radial velocity (middle),
and velocity dispersion (bottom) for the \emph{gEgSa05dir00} (black) and
\emph{gEgSb05dir00} (green) GalMer merger remnants. \label{gEgSa05dir00fig}}
\end{figure}

Structural properties of NGC~6340 can be reproduced by numerical
simulations. Exploring the GalMer database\footnote{http://galmer.obspm.fr/}
\citep{DiMatteo+08} we have found merger remnants having a three-component
brightness profiles qualitatively well resembling the one observed in
NGC~6340 but having larger sizes of all substructures suggesting the lower
masses of NGC~6340 progenitors than those of galaxies in the simulation. The
major equal-mass mergers between a non-rotating giant elliptical galaxy (E0)
and a spiral (Sa or Sb) on the co-planar prograde orbit of the configuration
\#5 (experiments \emph{gEgSa05dir00} and \emph{gEgSb05dir00}, see
\citealp{dMCMS07}), 3~Gyr after the beginning of the simulation result in a
formation of giant lenticular galaxies with large-scale rotating discs,
triaxial oblate nearly exponential bulges, and compact mass concentrations
in the circumnuclear regions. The kinematical misalignment is observed
between the outer disc and the exponential bulge of the remnants. The
metallicities of the central compact structures are strongly increased
during the strong circumnuclear starburst events induced by the interaction.
The gravitational softening of 0.28~kpc did not allow us to study the
internal structure of this central mass concentration. Several other
equal-mass mergers of spirals and ellipticals also result in remnants having
similar three-component density profiles. However, only the remnants of the
\emph{gEgSa05dir00} and \emph{gEgSb05dir00} encounters exhibits high degree of rotation in
the outer disc with $v_c / \sigma > 1$ (Fig.~\ref{gEgSa05dir00fig}).

Somewhat low metallicity of the large-scale disc of NGC~6340 favours the
major merger scenario: an equal mass merger as the simulated one would keep
a metallicity of the stars in the outer disc from the
intermediate-luminosity spiral galaxy having participated in the event,
therefore suggesting that the value we observe now should be slightly below
the one expected for an early-type galaxy with a mass of NGC~6340.

An interesting conclusion of our study is that the inner compact pseudo-bulge
has been formed as a result of a merger, and not through secular evolution,
one of the classical scenario to create pseudo-bulges
\citep{KK04}.

The interpretation of past events in NGC~6340 is not unique, apart from the
likely past merger. Indeed, the more recent gas accretion could be due to
tidal debris of the main merger, which take several Gyr to fall back. It
could also come from a more recent accretion of a small companion, that
could have created the shells, by distributing all its stars in the outer
disk. Alternatively, the gas accretion can come from the cosmic filaments,
which are not completely heated, since the galaxy is in a small group and
not in a cluster. 

The gas from filaments is expected to  become first very metal-poor ISM.
However, it might rapidly be enriched by subsequent star formation. We
cannot measure the gas metallicity directly with our data, however, the
presence of noticeable extinction and strong dust lanes does not favour the
idea of a low-metallicity ISM in the disc.

\begin{acknowledgements} 

Authors are grateful to the time allocation committee of the 6-m telescope
for providing observing time and to the SAO RAS staff members S.~Kaysin and
A.~Burenkov for the support of the service mode observations. AN and AZ
acknowledge the support of the Russian Foundation for Basic Research
(projects 07-02-0079 and 08-02-01323). AN thanks EARA ETN, Florence Durret
(IAP) and the LUTH laboratory of the Paris Observatory for funding her stay
in Paris. IC acknowledges additional support from the RFBR grant
07-02-00229-a. Special thanks to Olga Sil'chenko for useful discussions and
comments and to our anonymous referee for constructive advices which helped
us to improve the scientific discussion.

This research has made use of SAOImage DS9, developed by Smithsonian
Astrophysical Observatory; Aladin developed by the Centre de Donn\'ees
Astronomiques de Strasbourg; ``exploresdss'' script by G.~Mamon. Funding for
the SDSS and SDSS-II has been provided by the Alfred P. Sloan Foundation,
the Participating Institutions, the National Science Foundation, the U.S.
Department of Energy, the National Aeronautics and Space Administration, the
Japanese Monbukagakusho, the Max Planck Society, and the Higher Education
Funding Council for England. The SDSS Web Site is http://www.sdss.org/.

This work is based in part on archival data obtained with the Spitzer
Space Telescope, which is operated by the Jet Propulsion Laboratory,
California Institute of Technology under a contract with NASA.
\end{acknowledgements}

\bibliographystyle{aa}
\bibliography{ngc6340}


\end{document}